\documentclass[superscriptaddress,twocolumn,preprintnumbers,amsmath,amssymb]{revtex4-1}
\usepackage{graphicx}% Include figure files
\usepackage{dcolumn}% Align table columns on decimal point
\usepackage{booktabs,bm,color,braket}
\renewcommand{\theequation}{\arabic{equation}}
\usepackage{natbib}
\begin{document}

\title{Switchable Magnetic Bulk Photovoltaic Effect in the Two-Dimensional Magnet CrI$_3$}
\author{Yang Zhang}
\affiliation{Max Planck Institute for Chemical Physics of Solids, 01187 Dresden, Germany}
\affiliation{Leibniz Institute for Solid State and Materials Research, 01069 Dresden, Germany}
\affiliation{Department of Physics, Massachusetts Institute of Technology, Cambridge, Massachusetts 02139, USA}
\author{Tobias Holder}
\affiliation{Department of Condensed Matter Physics, Weizmann Institute of Science, Rehovot 7610001, Israel}
\author{Hiroaki Ishizuka}
\affiliation{Department of Applied Physics, University of Tokyo, Tokyo 113-8656, Japan}
\author{Fernando de Juan}
\affiliation{Donostia International Physics Center, P. Manuel de Lardizabal 4, 20018 Donostia-San Sebastian, Spain}
\affiliation{IKERBASQUE, Basque Foundation for Science, Maria Diaz de Haro 3, 48013 Bilbao, Spain}
\author{Naoto Nagaosa}
\affiliation{RIKEN Center for Emergent Matter Science (CEMS), Wako 351-0198, Japan}
\affiliation{Department of Applied Physics and Quantum Phase Electronics Center (QPEC), University of Tokyo, Tokyo 113-8656, Japan}
\author{Claudia Felser}
\affiliation{Max Planck Institute for Chemical Physics of Solids, 01187 Dresden, Germany}
\author{Binghai Yan $^\ast$}
\affiliation{Department of Condensed Matter Physics, Weizmann Institute of Science, Rehovot 7610001, Israel \\
Email: binghai.yan@weizmann.ac.il}

\begin{abstract}
The bulk photovoltaic effect (BPVE) rectifies light into the dc current in a single-phase material and attracts the interest to design high-efficiency solar cells beyond the p–n junction paradigm. Because it is a hot electron effect, the BPVE surpasses the thermodynamic Shockley-Queisser limit to generate above-band-gap photovoltage. While the guiding principle for BPVE materials is to break the crystal centrosymmetry, here we propose a magnetic photogalvanic effect (MPGE) that introduces the magnetism as a key ingredient and induces a giant BPVE. The MPGE emerges from the magnetism-induced asymmetry of the carrier velocity in the band structure. We demonstrate the MPGE in a layered magnetic insulator CrI3, with much larger photoconductivity than any previously reported results. The photocurrent can be reversed and switched by controllable magnetic transitions. Our work paves a pathway to search for magnetic photovoltaic materials and to design switchable devices combining magnetic, electronic, and optical functionalities.
\end{abstract}
\maketitle

\section*{Introduction}

Under strong light irradiation, a homogeneous non-centrosymmetric material can rectify  light into a dc current, called the bulk photovoltaic effect (BPVE)~\cite{Belinicher1980,von1981theory,Sturman1992,Sipe2000,Fridkin2001,young2012first,morimoto2016topological}. The induced open-circuit voltages can be much larger than the band gap. Thus, the BPVE displayed a promising potential in the solar energy conversion in ferroelectric perovskites~\cite{Choi2009,Yang2010,Grinberg2013} and  triggered the interest for the application of solar cells~\cite{Spanier2016} beyond the $p$--$n$  junction design and more recently also the application of photodetectors~\cite{Daranciang2012}.
It is believed that the shift current is dominant mechanism for the BPVE~\cite{von1981theory,Sipe2000,Tan2016}.
Shift current refers to real-space shift of conduction and valence Bloch electrons upon photoexcitation by a topological quantity, the Berry phase\cite{Xiao2010}. Therefore, ferroelectrics that exhibit intrinsic charge polarization are a promising direction in the search of BPVE materials. Furthermore, topological Weyl semimetals\cite{Yan2017,Armitage2017} have recently been investigated for promising BPVE \cite{wu2016giant,ma2017direct,chan2017photocurrents,deJuan2017quantized,zhang2018photogalvanic,zhang2018berry,osterhoudt2017colossal}, due to the large Berry phase in the band structure.

The essential requirement of the BPVE is inversion symmetry ($\mathcal{P}$) breaking. Thus, the present guiding principle to design BPVE materials is to break the \textit{crystal} centrosymmetry and sometimes to induce a strong charge polarization\cite{cook2017,nakamura2017shift,rangel2017,yang2018flexo}.
The lattice asymmetry or the polarization, however, is not the necessary condition for the inversion symmetry breaking. In this work, we show  a large photogalvanic effect by a magnetic ordering which breaks ${\cal P}$ but preserves the parity-time symmetry ($\mathcal{PT}$, where $\mathcal{T}$ represents the time-reversal symmetry); therefore, no polarization exists. This phenomenon, called magnetic photogalvanic effect (MPGE), can generate a photocurrent even upon the linearly polarized light. But it cannot be described by the shift current that applies to nonmagnetic systems. The MPGE is an intrinsic current response from the band-structure topology and distinct from the previously reported spin-galvanic effect~\cite{Ganichev2002} and magneto-gyrotropic photogalvanic effects~\cite{Ganichev2006} in semiconductor quantum wells (see Ref.~\onlinecite{Belkov2008} for review), which are driven by an external magnetic field, and also the spin BPVE that generates a spin current ~\cite{Ganichev2001,Young2013}.

The light excitation is known to generate an electron and hole pair in the solid. The velocity difference between the excited electron and hole may lead to a dc current.  However, such a current usually vanishes because the velocity reverses its direction from $\mathbf{k}$ to $\mathbf{-k}$ in the momentum space. Such a symmetry in the band structure (see Fig. 1) is induced by either $\mathcal{P}$ or $\mathcal{T}$. So we can realize this photocurrent by breaking both $\mathcal{P}$ and $\mathcal{T}$ to avoid the velocity cancellation in the band structure, which is the core of the MPGE proposed.

We demonstrate the MPGE in a newly discovered two-dimensional ferromagnetic insulator, CrI$_3$ ~\cite{huang2017b,mcguire2015coupling}. In the antiferromagnetic (AFM) phase of a CrI$_3$ bilayer~\cite{huang2017b,song2018giant}, both $\mathcal{P}$ and $\mathcal{T}$ are broken while the combined symmetry $\mathcal{PT}$ is preserved. We find that a giant dc photocurrent emerges in the visible light window. Because the $\mathcal{PT}$ symmetry forces the Berry phase to vanish, such a photocurrent is distinct from the shift current.
The photoconductivity is sourced from the resonant optical transition in the asymmetric band structure.
We found that the spin-orbit coupling (SOC) determines the extent of the momentum-inversion symmetry breaking and thus scales the amplitude of MPGE.
The magnetic phase transition, which is controllable as demonstrated in recent experiments ~\cite{Jiang2018,klein2018probing,song2018giant,jiang2018controlling,wang2018very}, can be utilized to control the direction and amplitude of the induced current. When reversing all spins in the AFM phase, we can switch the current direction. When switching from the AFM phase to the ferromagnetic (FM) phase, we can completely turn off the current by recovering the spatial inversion. Therefore, we realize three states with different light-matter responses -- positive, zero and negative currents (as illustrated in Fig. 1e). Thus, the MPGE provides a new pathway to control the light-matter interaction by magnetism and to design optical storage/switch devices. Our findings can be generalized to multilayers, the bulk system and other magnetic materials.

\section*{Results}

\subsection*{Symmetry of 2D magnetic insulator CrI$_3$}
The recent discovery of 2D van der Waals magnetic insulators, such as CrI$_3$, brings fascinating opportunities to design 2D magnetic devices. Bulk CrI$_3$ is an FM insulator~\cite{mcguire2015coupling}. The FM coupling is preserved down to the monolayer limit~\cite{huang2017b,Seyler2018}. In the bilayer and few-layer thickness, the interlayer coupling can be switched between AFM and FM by either an external magnetic field~\cite{klein2018probing} or electric gating ~\cite{jiang2018controlling,Jiang2018}, giving rise to a giant magnetoresistance effect~\cite{song2018giant,wang2018very}.
The atomic crystal of CrI$_3$ exhibits inversion symmetry with two inversion centers, one inside the monolayer and the other in between neighboring monolayers. The AFM order reduces the crystal symmetry by removing the interlayer inversion center. Therefore, an AFM bilayer breaks the inversion symmetry, although an FM bilayer does not.  Another important feature is that the AFM ordered phase preserves $\mathcal{PT}$ symmetry while it breaks $\mathcal{P}$ and $\mathcal{T}$ independently; hence, no polarization exists in the ordered phase.
Therefore, the bilayer CrI$_3$ (see Fig. 1) is an ideal system to examine the photocurrent response, where the inversion symmetry is solely broken by the AFM order instead of the crystal structure.
%Although both $\mathcal{P}$ and $\mathcal{T}$ are broken, their combination $\mathcal{PT}$, also called parity-time symmetry, presents a new symmetry in the AFM bilayer.
In the following, the $\mathcal{PT}$ symmetry is found to be crucial to determine the mechanism of the photocurrent response.

\begin{figure*}[htbp]
\begin{center}
\includegraphics[width=0.9\textwidth]{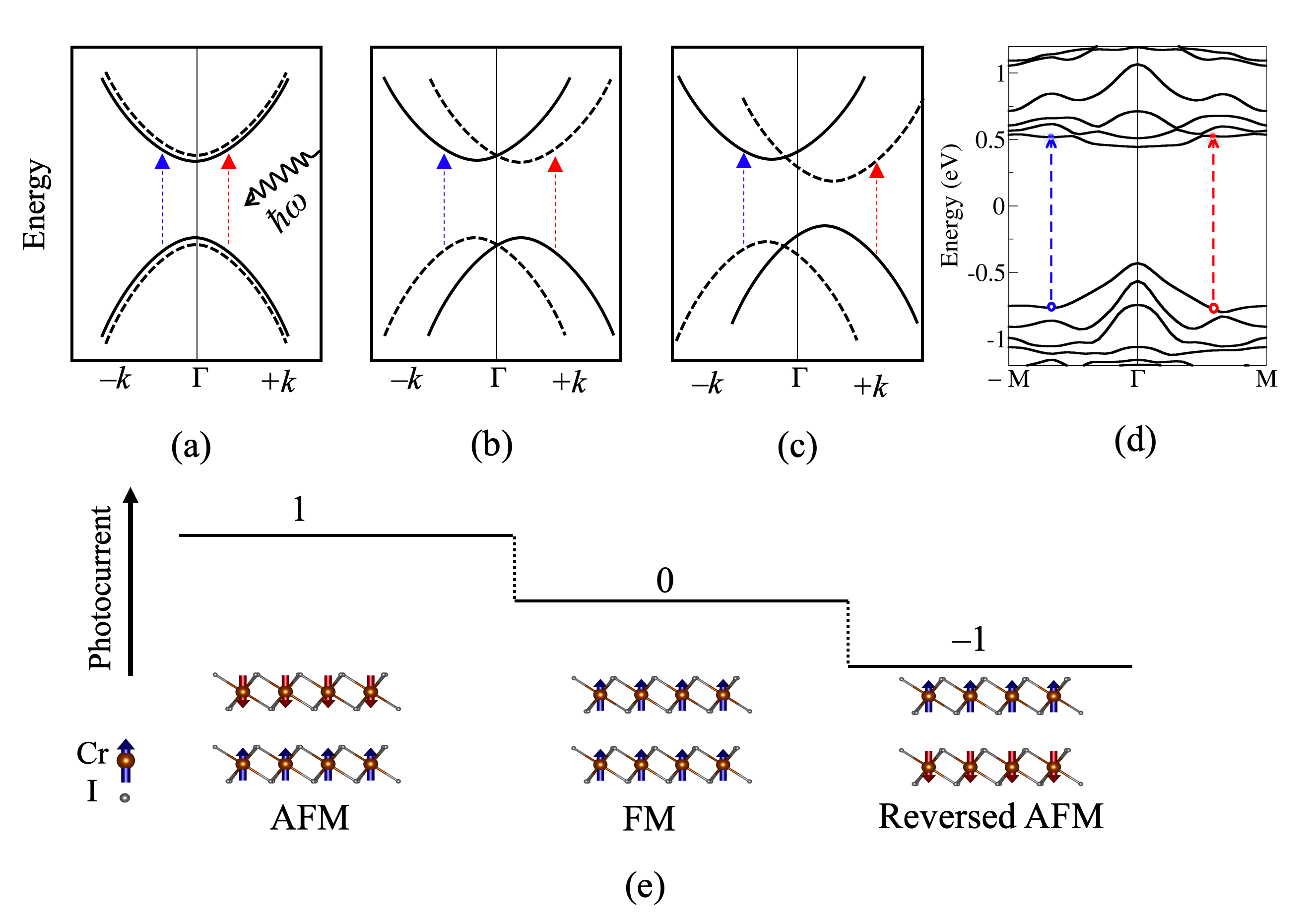}
\end{center}
\caption{\textbf{Band structure symmetry-breaking and magnetic structures of the bilayer CrI$_3$.} Schematics of band structures (a) with both inversion symmetry ($\mathcal{P}$) and the time-reversal symmetry ($\mathcal{T}$), (b) with only $\mathcal{T}$ but $\mathcal{P}$-breaking, and (c) with both $\mathcal{P}$- and $\mathcal{T}$-breaking.
For both (a) and (b), the light excitation ($\hbar \omega$) at $\mathbf{+k}$ and $\mathbf{-k}$ is symmetric to each other. However, such a symmetry is broken in (c). As a consequence, excited electrons at $\mathbf{+k}$ and $\mathbf{-k}$ do not cancel each other in velocity, giving rise to a dc photocurrent.
(d) The band structure of antiferromagnetic(AFM) bilayer CrI$_3$.
Here both $\mathcal{P}$- and $\mathcal{T}$ are broken as the case of (c), violating the  $\mathbf{k}$ to $\mathbf{-k}$ symmetry.
The spin-orbit coupling is included. The Fermi energy is shifted to zero. (e) The AFM, ferromagnetic (FM) and reversed AFM phases display three distinct responses to a linearly polarized light -- positive current state(1), zero current state (0) and negative current state(-1). }
\end{figure*}

We show the band structures of the bilayer calculated by the density-functional theory (DFT) including the spin-orbit coupling (SOC) in Fig. 1. All bands are doubly degenerate which is protected by the $\mathcal{PT}$ symmetry in the AFM phase. In contrast, such degeneracy is lifted in the FM phase (Supplementary Fig. 1).
For the AFM phase, an important feature is the breaking of the $\mathbf{k}$ to $\mathbf{-k}$ symmetry. In a simple consequence, excited electrons (holes) at $\mathbf{k}$ and $\mathbf{-k}$ by the optical excitation ($\hbar \omega$) exhibit uncompensated velocities and lead to a nonzero dc current. In addition, the energy gap is slightly lower than the experimental value, which can be attributed to the known gap underestimation of DFT.

\subsection*{Photocurrent of bilayer CrI$_3$}
In general, the photocurrent is a nonlinear effect. In this work, we provide a proper formalism (Equation~\eqref{second-kubo}) to describe $\mathcal{T}$-breaking photocurrent by the second-order response theory (see Sec.\ref{discussion}). We first evaluate the photocurrent response of the bilayer CrI$_3$ using the general formalism based on Bloch wave functions of the realistic material. Then we prove that this formalism can be reduced to a simple form of the resonant optical transition (Equation~\eqref{transition}) in the presence of the $\mathcal{PT}$ symmetry.

Under the irridation of the linearly polarized light, the FM bilayer exhibits vanishing photocurrent due to inversion symmetry. However, the AFM phase displays a large photocurrent conductivity (more than 200 $\mu A V^{-2}$ for a relaxation time $\tau \approx 0.4$ ps, i.e. $\hbar / \tau = 1$ meV) in the visible-light range (see Fig. 2). This value is higher than that of many known BPVE materials reported so far~\cite{young2012first,young2012BFO,zheng2014,brehm2014,cook2017,rangel2017,sotome2019spectral,Fei2018,Chan2019}. We note that the value of the photocurrent is proportional to the relaxation time (see Supplementary Fig. 3). Here we choose the $\tau$ value according to the magnitude of experimental reports on other transition metal dichalcogenides\cite{Kozawa2014,Sohier2018}. We note that the photocurrent reverses its direction when reversing all spins in the AFM phase. Therefore, the FM, AFM and reversed AFM represent three photocurrent states, 0 (no current), 1 (positive current) and -1 (negative current), respectively, as illustrated in Fig. 1. In addition, for circularly polarized light, we do not observe any photocurrent for both AFM and FM bilayers, which is constrained by the 2D point group symmetry.

\begin{figure*}[htbp]
\begin{center}
\includegraphics[width=0.6\textwidth]{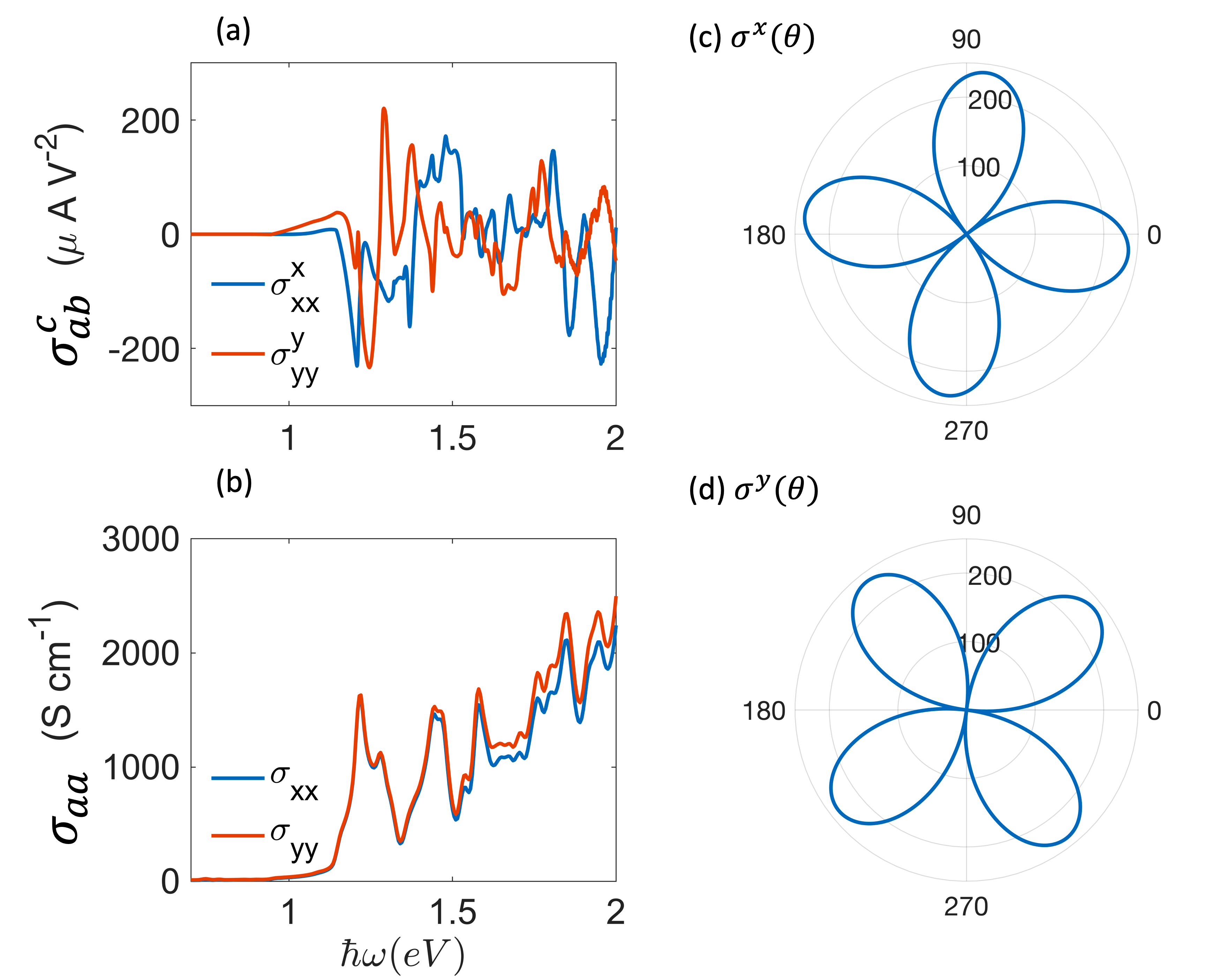}
\end{center}
\caption{\textbf{Calculated photoconductivity in response to the linearly polarized light.} (a) The photon energy ($\hbar \omega$) dependence of $\sigma^x_{xx}$ and $\sigma^y_{yy}$. (b) The linear-response optical conductivity  $\sigma_{xx}$ and $\sigma_{yy}$. (c)-(d) The angle dependent photoconductivity $\sigma^x (\theta)$,$\sigma^y (\theta)$ [cf. Eq~(\eqref{compactcurrent})] in the same unit as (a) for $\hbar \omega = 1.2 $ eV and $\hbar / \tau = 1$ meV. $x$ and $y$ are the directions of the current and $\theta$ is the angle between the electric field of light and the $x$-axis. }
\end{figure*}

\section*{Discussion}
\label{discussion}
We describe the BPVE response by the general second-order Kubo formalism  ~\cite{von1981theory} and then drive the MPGE in the condition of $\mathcal{T}$-breaking. The general theory accounts for the steady-state short-circuit photocurrent with the relaxation time approximation. The conductivity $\sigma^c_{ab}$ ($a,b,c=x,y$ in 2D) represents the photocurrent $J^c$ generated by the dipole electrical field of light, $\mathbf{E}=(E_a(\omega),E_b(\omega),0)$,
% \begin{subequations}
\begin{alignat}{2}
\label{Jc}
J^c & =  \sum\limits_{ab} \sigma^c_{ab}E_a^{\ast}(\omega)E_b(\omega) \\
\label{second-kubo}
    & =  \sum\limits_{ab} \frac{\pi e^3}{ \omega^2} Re \bigg\{
  \sum_{l,m,n}^{\Omega=\pm \omega} \int_{BZ} \frac{d\mathbf{k}}{(2\pi)^d} f_{ln}
   \frac{v^a_{nl} v^b_{lm} v^c_{mn}}
   {(E_{nm}-i\hbar/\tau)(E_{nl} - \hbar \Omega- i\hbar/\tau)}  E_a^{\ast}(\omega)E_b(\omega) \bigg\},
\end{alignat}
% \end{subequations}
where $v^a_{nl}=   \bra{n, \mathbf{k}} \hat{v}^a \ket{l, \mathbf{k}} $, $\hat{v}^a$ is the velocity operator, $\tau$ is the relaxation time, $d$ is the system dimension,
$f_{ln}=f(E_l) - f(E_n)$, $f(E_n)$ is the Fermi-Dirac distribution,  $E_{nm}=E_n - E_m$, $E_n\equiv E_n(\textbf{k})$ and $\ket{n,\textbf{k}}$ are energy and wave functions, respectively, at $\textbf{k}$ for the $n$th band.
%We note that $\phi_{ab}$ is the phase shift between $E_a$ and $E_b$ and thus $\phi_{ab}=0$ and $\phi_{ab}=\pi/2$ represent the linearly and circularly ($a \neq b$) polarized light, respectively.

The photocurrent conductivity matrix $\sigma_{ab}^c$ is shaped by the symmetry of the AFM bilayer. The three-fold rotational symmetry and $\mathcal{PT}$ lead to only six nonzero tensor elements, $\sigma^x_{xx} = -\sigma^x_{yy}=-\sigma^y_{xy}=-\sigma^y_{yx}$ and $\sigma^y_{yy} = -\sigma^y_{xx}=-\sigma^x_{yx}=-\sigma^x_{xy}$. Here only two of them (such as $\sigma^x_{xx}$ and $\sigma^y_{yy}$) are independent, since $x$ and $y$ directions are not equivalent in a hexagonal lattice.
For a linearly polarized light $\mathbf{E}=(\cos \theta,\sin \theta,0) E_0 \cos (\omega t)$, according to Equation~\eqref{Jc} the photocurrent along the $x$ and $y$ direction is
\begin{align}
J^x&= \sigma^x (\theta) E_0^2 = [\sigma^x_{xx} \cos (2\theta) - \sigma^y_{yy} \sin (2\theta)] E_0^2\notag\\
J^y&=  \sigma^y (\theta) E_0^2 = [-\sigma^y_{yy} \cos (2\theta) - \sigma^x_{xx} \sin (2\theta)] E_0^2.
\label{compactcurrent}
\end{align}
The photocurrent is sensitive to the polarization direction $\theta$, as we show in Figs. 2c-2d. Following Equation \eqref{compactcurrent}, if $\sigma^y_{yy}$ is zero, the maxima of $\sigma^x(\theta)$ is located at $\theta =0,180 ^\circ$. Because $\sigma^y_{yy}$ is generically nonzero at a given frequency, however, the maxima of $\sigma^x$ shift away from $\theta =0,180 ^\circ$ in Figs. 2c. Such an anisotropy is usually measured to deduce the conductivity tensor elements in experiment.

In contrast, for the circularly polarized light, the photocurrent is forced to vanish by the 2D point group symmetry ($C_3$) despite that it may appear in a 3D material.
In other words, $J^x=\sum\limits_{ab} \sigma^x_{ab}E_a^{\ast}E_b=\sigma^x_{xx}E_x^{\ast}E_x + \sigma^x_{yy}E_y^{\ast}E_y + \sigma^x_{xy}E_x^{\ast}E_y + \sigma^x_{yx}E_y^{\ast}E_x  = 0$ for a circular polarized light with $E_y = i E_x $. Such a dramatic distinction between light polarization provides a simple, useful hallmark to verify these results on 2D materials.
We will focus our discussions mainly on the linearly polarized light in the following.

The general formalism (Equation~{second-kubo}) can be simplified when special symmetries appear. The BPVE arises as a consequence of the three-band ($l,m,n$) interference in the optical excitation. (i) The response function vanishes to zero if the inversion $\mathcal{P}$ exists, because the numerator $N_{lmn}(\mathbf{k})=v^a_{nl} v^b_{lm} v^c_{mn}$, the product of three matrix elements, is odd to $\mathbf{k}$. (ii) If $\mathcal{P}$ is broken but $\mathcal{T}$ appears, $N_{lmn}(\mathbf{k})=-N_{lmn}(-\mathbf{k})^{\ast}$. Then only the imaginary part of $N_{lmn}(\mathbf{k})$ contributes nonzero values to the photocurrent. Therefore, the linearly and circularly polarized lights are related to the imaginary and real parts of the energy denominator, respectively, in Equation~\eqref{second-kubo}.
For a $\mathcal{T}$-symmetric insulator, Equation~\eqref{second-kubo} can be simplified to the shift current and injection current formalisms~\cite{von1981theory,aversa1995,Sipe2000} for the linearly and circularly polarized lights, respectively.
In non-magnetic systems, this implies that the circular photocurrent scales linearly with the scattering rate, while the linear photocurrent is independent of the scattering rate.
Here, the velocity matrix is commonly transformed to the length gauge~\cite{aversa1995}.
%, because valence and conduction wave functions are smooth in the momentum space.
%For a gapless system with crossing bands (e.g. a Weyl semimetal) ~\cite{von1981theory,zhang2018photogalvanic}, however, the general formula Equation~\eqref{second-kubo} should be employed.

For the bilayer AFM CrI$_3$ that respects $\mathcal{PT}$ but breaks $\mathcal{P}$ and $\mathcal{T}$ independently, the response function exhibits a unique symmetry. Because $\mathcal{PT}$ requires $N_{lmn}(\mathbf{k}) =  N_{l^\prime m^\prime n^\prime}^{\ast}(\mathbf{k})$, the numerator [$N_{lmn}(\mathbf{k})+N_{l^\prime m^\prime n^\prime}^{\ast}(\mathbf{k})$] in Equation~\eqref{second-kubo} is always real, where $l^\prime ,m^\prime ,n^\prime$ are $\mathcal{PT}$ partners (degenerate in energy) of $l, m, n$ , respectively. Therefore, the linearly and circularly polarized lights are related to the real and imaginary parts of the energy denominator, respectively, opposite to the $\mathcal{T}$-symmetric case.

%In the $\mathcal{PT}$-symmetric system, the Abelian Berry phase vanishes while the non-Abelian Berry phase still appears. Because both Abelian and non-Abelian parts are related to the imaginary part of $N_{lmn}$, they will not contribute a photocurrent under the $\mathcal{PT}$ symmetry.

To establish the intuitive correlation between the band structure and the BPVE, we decompose Equation~\eqref{second-kubo} into two-band and three-band process~\cite{von1981theory,zhang2018photogalvanic}. The former corresponds to a direct resonant transition from $\ket{l}$ to $\ket{n}$, where $n=m$.
By considering $(E_n-E_m-i \hbar/\tau)^{-1}= -i \tau / \hbar $ in Equation~\eqref{second-kubo}, the response function can be derived,
\begin{equation} \label{PT-injection}
  \sigma^c_{ab} = -\frac{\pi e^3 \tau}{ \omega^2 \hbar}
  \sum_{l,n; \Omega=\pm \omega} \int_{BZ} \frac{d\mathbf{k}}{(2\pi)^d} f_{ln}
   \frac{1}{2}\{v^a_{nl}, v^b_{ln}\} v^c_{nn}
   \delta(E_{nl} - \hbar \Omega),
 \end{equation}
where $ \frac{1}{2}\{v^a_{nl}, v^b_{ln}\} =\frac{1}{2}(v_{nl}^a v_{ln}^b + v_{ln}^a v_{nl}^b) \equiv Re(v_{nl}^a v_{ln}^b)$ and the $\delta$-function is derived from the imaginary part of $(E_{nl} - \hbar \Omega-i \hbar / \tau)^{-1}$.
Such two-band photocurrent is proportional to the relaxation time $\tau$ and decided by the resonant transition according the selection rule $ E_{nl} =\pm \hbar \omega$.
We further transform Equation~\eqref{PT-injection} to the length gauge,
\begin{equation}
\label{injection}
\sigma^c_{ab} = \frac{\pi e^3 \tau}{ \hbar}
  \sum_{l,n} \int_{BZ} \frac{d\mathbf{k}}{(2\pi)^d} f_{ln}
   \frac{1}{2}\{r_{nl}^a, r_{ln}^b\}  \Delta_{ln}^{c}
   \delta(E_{nl} - \hbar \omega),
\end{equation}
 where $r^a_{nl}=i \bra{n} \partial_{k_a} \ket{l}$ is the position matrix element and $v^a_{nl}= r^a_{nl} E_{nl}/\hbar \equiv r^a_{nl}\Omega$ if $E_{nl}-\hbar \Omega =0$,  and $\Delta^c_{ln}=v^c_{ll}-v^c_{nn}$.
 This  formula is very similar to the the known $\mathcal{T}$-symmetric injection
  current expression(Equation 56 in Ref.~\onlinecite{Sipe2000}), except that in Ref.~\onlinecite{Sipe2000} the injection current is integrated over the imaginary part of the position matrix $
 r_{nl}^a r_{ln}^b$ due to $\mathcal{T}$. In contrast, Equation~\eqref{injection} evaluates its real part due to $\mathcal{PT}$. Suppose a linearly polarized light with the polarization along $x$, Equation~\eqref{injection} looks more intuitive in the following form,
 \begin{equation}
\label{transition}
\sigma^c_{xx} = \frac{\pi e^3 \tau}{ \hbar}
  \sum_{l,n} \int_{BZ} \frac{d\mathbf{k}}{(2\pi)^d} f_{ln}
   |r_{ln}^x|^2 \Delta_{ln}^{c} \delta(E_{nl} - \hbar \omega).
\end{equation}
It represents an excitation from $l$ to $n$ with dipole transition rate $|r_{ln}^x|^2$. The excited electron (hole) with finite velocity $v^c_{nn}$ ($v^c_{ll}$) induces a dc current. If $v^c_{nn}$ at $\mathbf{k}$ and $\mathbf{-k}$ cancel each other exactly when $\mathcal{P}$ or $\mathcal{T}$ exists, the photocurrent from Equation~\eqref{injection} or ~\eqref{transition} vanishes. Only when both $\mathcal{P}$ and $\mathcal{T}$ are broken, a nonzero photocurrent may exist.

The three-band contribution refers to $n \neq m$. We can exclude a trivial case that $\ket{n}$ and $\ket{m}$ are degenerate in energy for example protected by $\mathcal{PT}$, because $v^a_{nm}=0$ if so. When $E_n \neq E_m$,
the real part of the denominator in Equation~\eqref{second-kubo} is $\tau$-independent. Thus the three-band process generates a photocurrent that is robust against scattering, different from the two-band transition.
The three-band process evaluate $(E_{nm}-i \hbar/\tau)^{-1}$ in the denominator. Thus, the ratio between the three-band and two-band contributions are $(\hbar/\tau) / E_{nm}$ that is usually in the order of meV/eV. Then the three-band contribution is much smaller than the two-band one. In calculations, we indeed find that the photoconductivity is predominantly produced by the two-band transition (the three-band contribution is less than 0.5\%, See the supplementary Fig. 2).

\begin{figure*}[htbp]
\begin{center}
\includegraphics[width=1\textwidth]{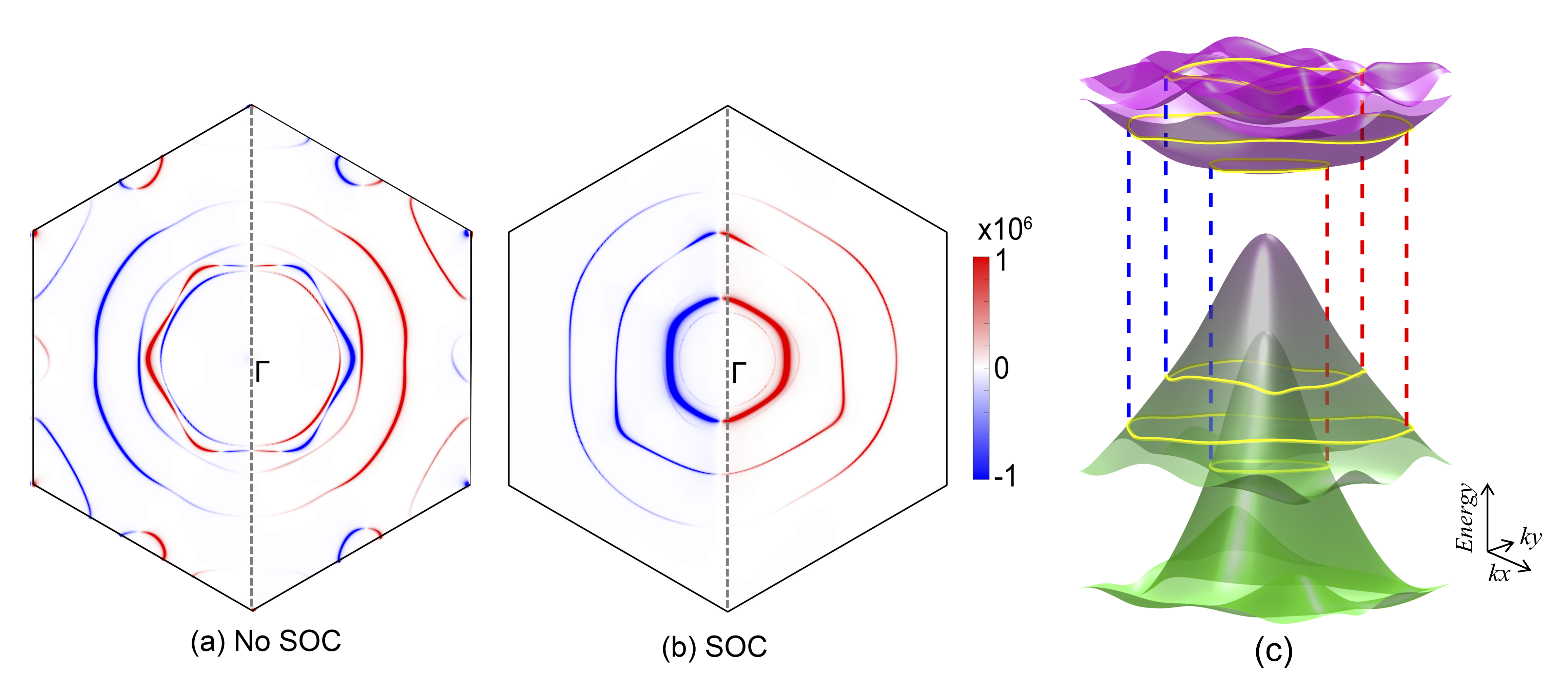}
\end{center}
\caption{\textbf{Distribution of the photoconductivity $\sigma^x_{xx}$ in the first Brillouin zone.} Without including the spin-orbit coupling (SOC), the $\mathbf{k}$ to $\mathbf{-k}$ symmetry appears while finite SOC breaks such a symmetry. The ring-like shape indicates the resonant optical transition between valence ($E_l$) and conduction ($E_n$) bands by the selection rule $E_n - E_l = \hbar \omega$ (1.2  eV). (c) Transitions from top two valence bands to the bottom two conduction bands. The yellow rings indicate the transition paths and correspond to the large-$\sigma^x_{xx}$-amplitude rings in (b).}
\end{figure*}

It is useful to find a simple indicator for the nonlinear photoconductivity. Because the photocurrent is proportional to the resonant transition rate, the imaginary dielectric function or the optical conductivity may provide such an indicator. In other words, the optical conductivity is the sum of all allowed transitions equally while the photoconductivity is to sum them with the velocity weight (see Equation~\eqref{transition}). As shown in Fig. 2c, the optical conductivity $\sigma_{xx}$ is strongly correlated to $\sigma^x_{xx}$. In addition, the photoconductivity indeed scales linearly to the relaxation time $\tau$ in our calculations (Supplementary Fig. 3).

Figure 3 shows the $\sigma^x_{xx}$ distribution in the momentum space for the AFM bilayer at $\hbar \omega = 1.2 $ eV. It is clear that the photoconductivity is contributed by resonant transition channels between the valence and conduction bands. Because of the breaking of both $\mathcal{P}$ and $\mathcal{T}$, these channels are not symmetric between $\mathbf{k}$ and $\mathbf{-k}$ anymore. We point out that SOC play a significant role here despite that it is less obvious from Equation~\eqref{transition}. The strength of SOC represents the amount of the $\mathbf{k}$ to $\mathbf{-k}$ symmetry-breaking in the band structure. When SOC is absent, the AFM band structure is still symmetric (see the supplementary Fig. 1). This is because of the spin rotation symmetry SU(2). Therefore, the net photocurrent vanishes, as shown in Fig. 3a. Finite SOC locks the the spin orientation with respect to the lattice and breaks the momentum-inversion symmetry (see Fig. 3b), resulting in a nonzero photocurrent. The SOC modifies the band structure by opening a gap at the band crossing points. So peaks of $\sigma^x_{xx}$ usually correspond to the transitions involving these anti-crossing gap regions.

In reality, the substrate or the gate may modify the bilayer electronic structure by breaking the crystal inversion between two layers. It is practical to investigate how robust the $\mathcal{PT}$-symmetry-induced photocurrent behaves in the presence of such perturbation. We apply an out-of-plane electric field $E$ to represent the perturbation. $E$ induces a potential drop between two layers and breaks the $\mathcal{PT}$ symmetry of the bilayer. We calculate the photoconductivity using Equation~\eqref{second-kubo}, where both the real and imaginary parts of $N_{lmn}(\mathbf{k})$ contribute. The photoconductivity remains relatively robust even for a large $E =$ 0.005 V/$a.u.$ ( 1 $a.u. = 0.53$ \AA) for the AFM bilayer (see the supplementary Fig. 4). On the other hand, the FM phase starts to exhibit nonzero photocurrent when $E$ breaks its inversion symmetry. At $E =$ 0.005 V/$a.u.$, the photoconductivity is of the same order as in the AFM case.

In addition, the MPGE can be easily generalized to multi-layers and other magnetic materials (such as  Cr$_2$Ge$_2$Te$_6$\cite{Gong2017} and VSe$_2$ ~\cite{Bonilla2018}).
Some magnetic orders of a trilayer CrI$_3$, for example, $\uparrow \uparrow \downarrow$ or $\uparrow \downarrow \downarrow$,  can break the inversion symmetry and generate the MPGE . Different from a bilayer, $\mathcal{PT}$ is broken here. Then the real and imaginary parts of $N_{lmn}(\mathbf{k})$ contribute to the photoconductivity in Equation~\eqref{second-kubo}. Magnetic orders such as $\uparrow \uparrow \uparrow$ and $\uparrow \downarrow \uparrow$ preserve the inversion symmetry and produce no photocurrent.
Further, hetero-structures of layered materials and twisted layers provide vast possibilities.
Beyond 2D materials, the MPGE may also exist in 3D systems, specially AFM materials with the $\mathcal{PT}$ symmetry and strong SOC, such as Cr$_2$O$_3$~\cite{Fiebig1994}, Mn$_2$Au~\cite{Barthem2013}, CuMnAs~\cite{Olejnik2017,Emmanouilidou2017}.

%\section*{Summary}

To summarize, we discover a magnetic photogalvanic effect to generate the photocurrent by breaking the momentum-inversion symmetry in the band structure. This mechanism induces a large photoconductivity in the AFM phase of the bilayer CrI$_3$ despite no electric polarization.
The photocurrent appears for the linearly polarized light but vanishes for the circularly polarized one in the 2D system.
It exhibits an injection-current-like feature and is
proportional to the relaxation time, the resonant transition rate and SOC. Tuning the magnetic structure is a sensitive handle to manipulate the photocurrent. Although the bilayer CrI$_3$ exhibits a record photoconductivity, it seems possible to design even better nonlinear magnetic materials, for example by considering stronger SOC and longer relaxation time. There are plenty of AFM and FM materials, such as magnetoelectric multiferroic compounds~\cite{fiebig2005revival,Spaldin2005}, waiting for exploration.

\section*{Method}
We obtain the DFT band structure and Bloch wave functions from the Full-Potential Local-Orbital program (\textsc{FPLO})~\cite{koepernik1999full} within the generalized gradient approximation~\cite{perdew1996}. By projecting the Bloch wave functions onto atomic-like (Cr-$d$ and I-$sp$) Wannier functions, we obtain a tight-binding Hamiltonian with sixty-eight bands that well reproduce the DFT band structure.  We employ this material specific tight-binding Hamiltonian for accurate evaluation of the photocurrent. We use $\hbar / \tau = 1$ meV in our calculations. For the integrals of Eq. 1b, the 2D Brillouin zone was sampled by a grid of $400 \times 400$. Increasing the grid  size to $960\times960$ varied the conductivity by less than 5\%.The spin-orbit coupling was included in a self-conssistent way for the DFT calculations. The FPLO band structure is well consistent with other DFT methods.

\section*{Data availability}
The data that support the findings of this study are available
from the corresponding author upon request.

\section*{Acknowledgments}
We thank Liang Wu, Markus Mittnenzweig, Adolfo G. Grushin and Liang Fu for helpful discussions. N.N. was supported by JST CREST Grant Number JPMJCR1874 and JPMJCR16F1, Japan, and JSPS KAKENHI Grant numbers 18H03676 and 26103006.
H.I. was supported by JSPS KAKENHI Grant Numbers JP18H04222, JP19K14649, and JP18H03676.
C.F. was funded by the DFG through SFB 1143 (project ID 247310070) and thanks the Wuerzburg-Dresden Cluster of Excellence on Complexity and Topology
in Quantum Matter - ct.qmat (EXC 2147, project ID 39085490).
B.Y. acknowledges the financial support by the Willner Family Leadership Institute for the Weizmann Institute of Science, the Benoziyo Endowment Fund for the Advancement of Science,  Ruth and Herman Albert Scholars Program for New Scientists, the European Research Council (ERC) under the European Union’s Horizon 2020 research and innovation programme (grant agreement No. 815869) and by the collaborative Max Planck Lab on Topological Materials.
% add Claudia's grants in next version

\section*{Author contributions}
B.Y. conceived the project. Y.Z. did the DFT and photocurrent calculations. Y.Z. and B.Y. performed the derivation of photocurrent response. Y.Z., T.H., F.d.J., H.I., and B.Y. analyzed the results. Y.Z. and B.Y. wrote the manuscript with the input from all other authors. N.N., C.F. and B.Y. supervised the project.

\section*{Competing  Interests}
The Authors declare no competing interests.

%\bibliography{ref}% Produces the bibliography via BibTeX.

\clearpage
% Add 'S' to the numbering inside the supplement
\renewcommand{\thesection}{S\arabic{section}}
\renewcommand{\thetable}{S\arabic{table}}
\renewcommand{\thefigure}{S\arabic{figure}}
\renewcommand{\theequation}{S\arabic{equation}}

\setcounter{section}{0}
\setcounter{figure}{0}
\setcounter{table}{0}
\setcounter{equation}{0}

%{\bf Supplemental Materials}

\setcounter{figure}{0}

\begin{figure*}[h]
\includegraphics[width=0.8\textwidth]{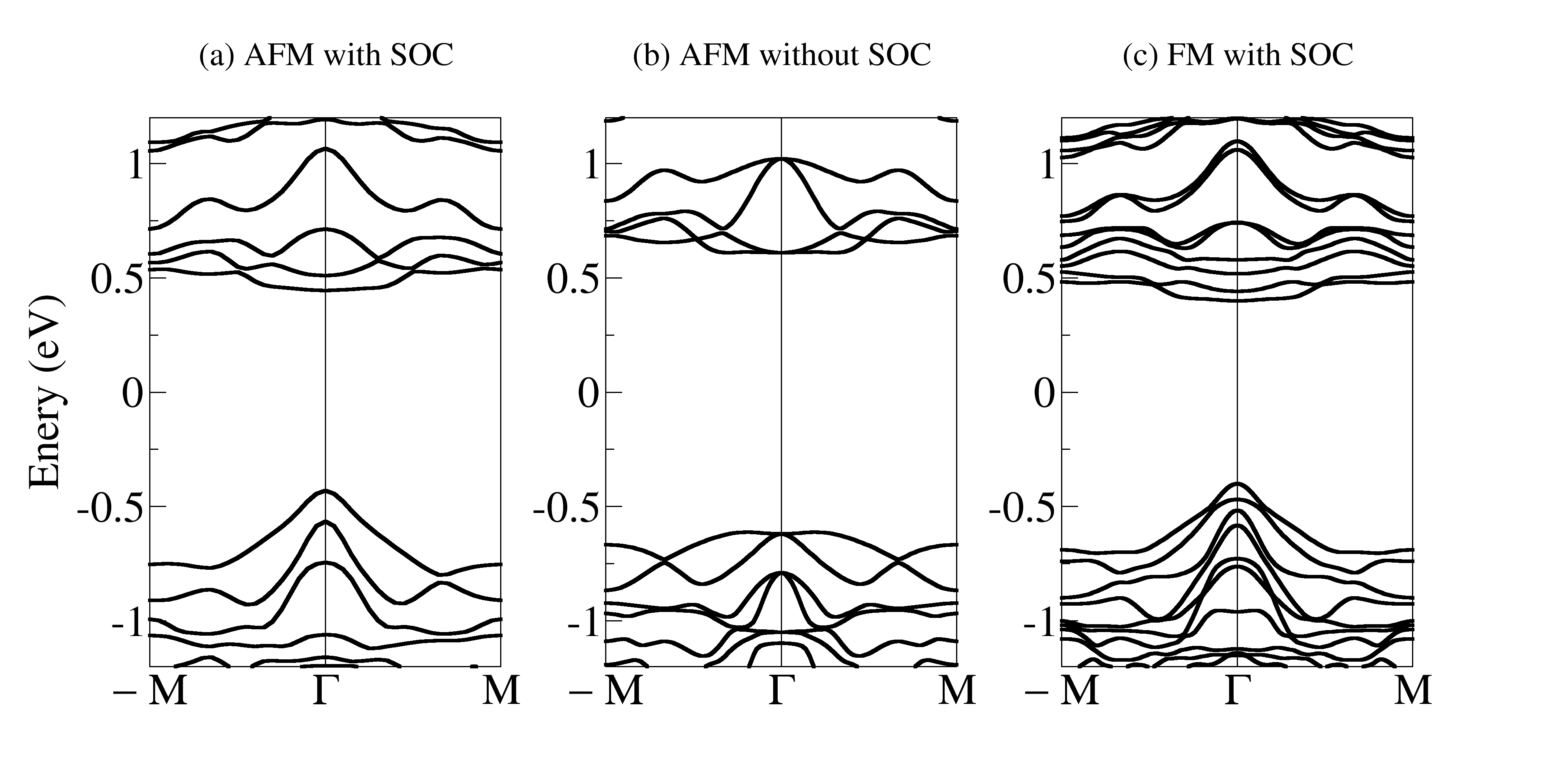}
\caption{Band structure of the bilayer CrI$_3$ for the (a) AFM phase with SOC, (b) AFM without including SOC and (c) FM phase with SOC. In both (b) and (c), the $\mathbf{k}$ to $\mathbf{-k}$ symmetry is preserved in the band structure. In contrast, momentum inversion symmetry is broken in (a).
}
\label{SM-band}
\end{figure*}

\begin{figure*}[h]
\includegraphics[width=0.8\textwidth]{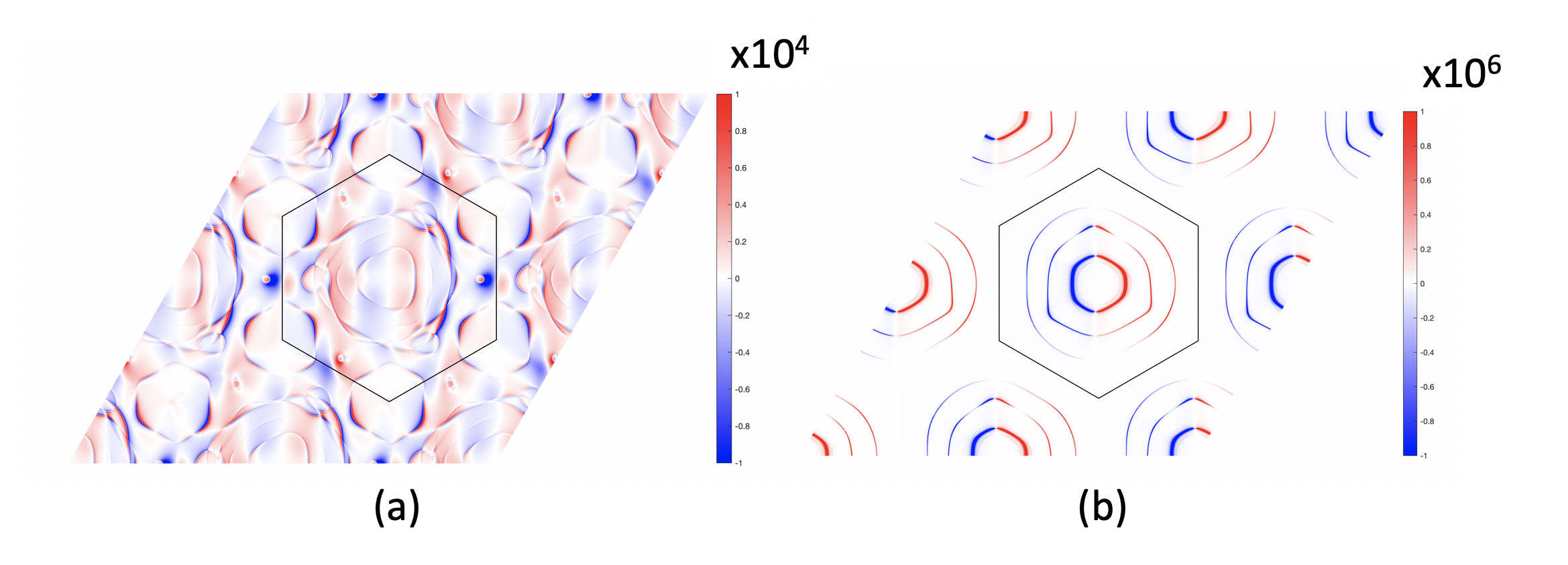}
\caption{The photoconductivity contributions from (a) the three-band and (b) two-band processes. Distribution of the photoconductivity $\sigma^x_{xx}$ ($\hbar \omega =1.2$  eV) in the first Brillouin zone, the hexagonal area. Note that the colorbar of the three-band contribution is two orders in magnitude smaller than the two-band one. The momentum inversion symmetry is broken in both (a) and (b). The three-band distribution is relatively uniform in the Brillouin zone, because it has no energy selection rule.
}
\label{SM-3band}
\end{figure*}

\begin{figure*}
\includegraphics[width=0.9\textwidth]{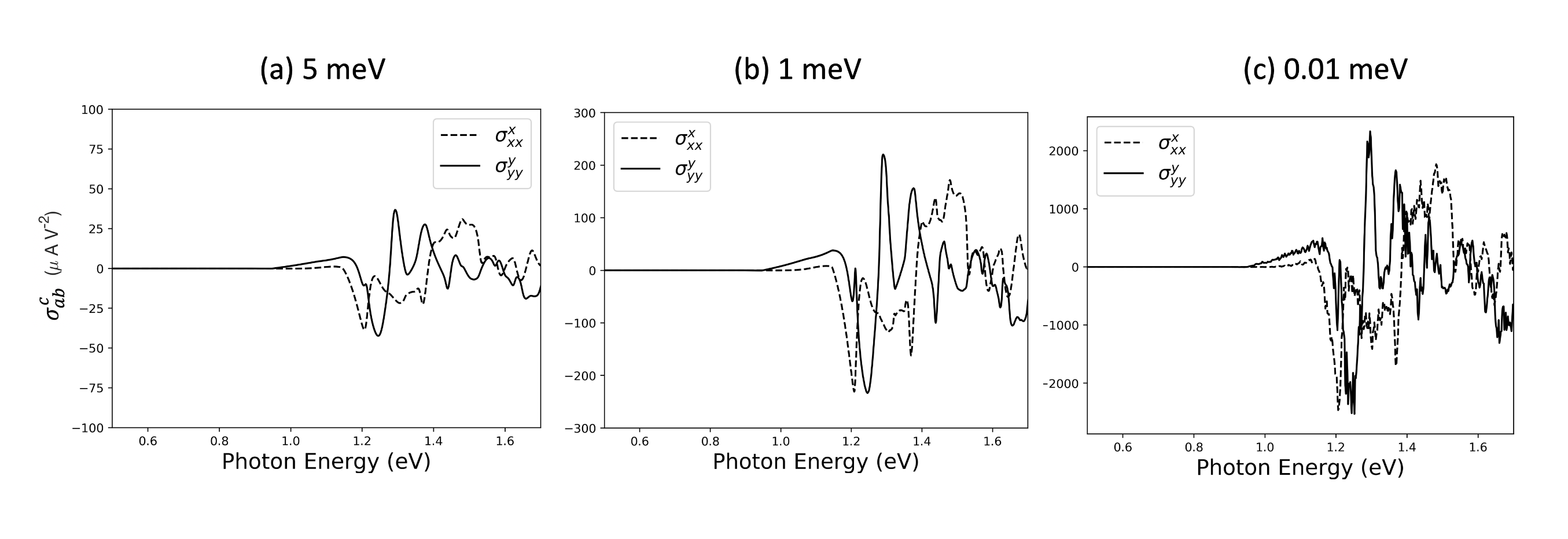}
\caption{The photoconductivity for different relaxation time (a) $\hbar / \tau = 5$ meV, (b) 1 meV (same as Fig. 2a) and (c) 0.01 meV. The photoconductivity scales linearly with $\tau$.}
\label{SM-tau}
\end{figure*}

\begin{figure*}
\begin{center}
\includegraphics[width=0.6\textwidth]{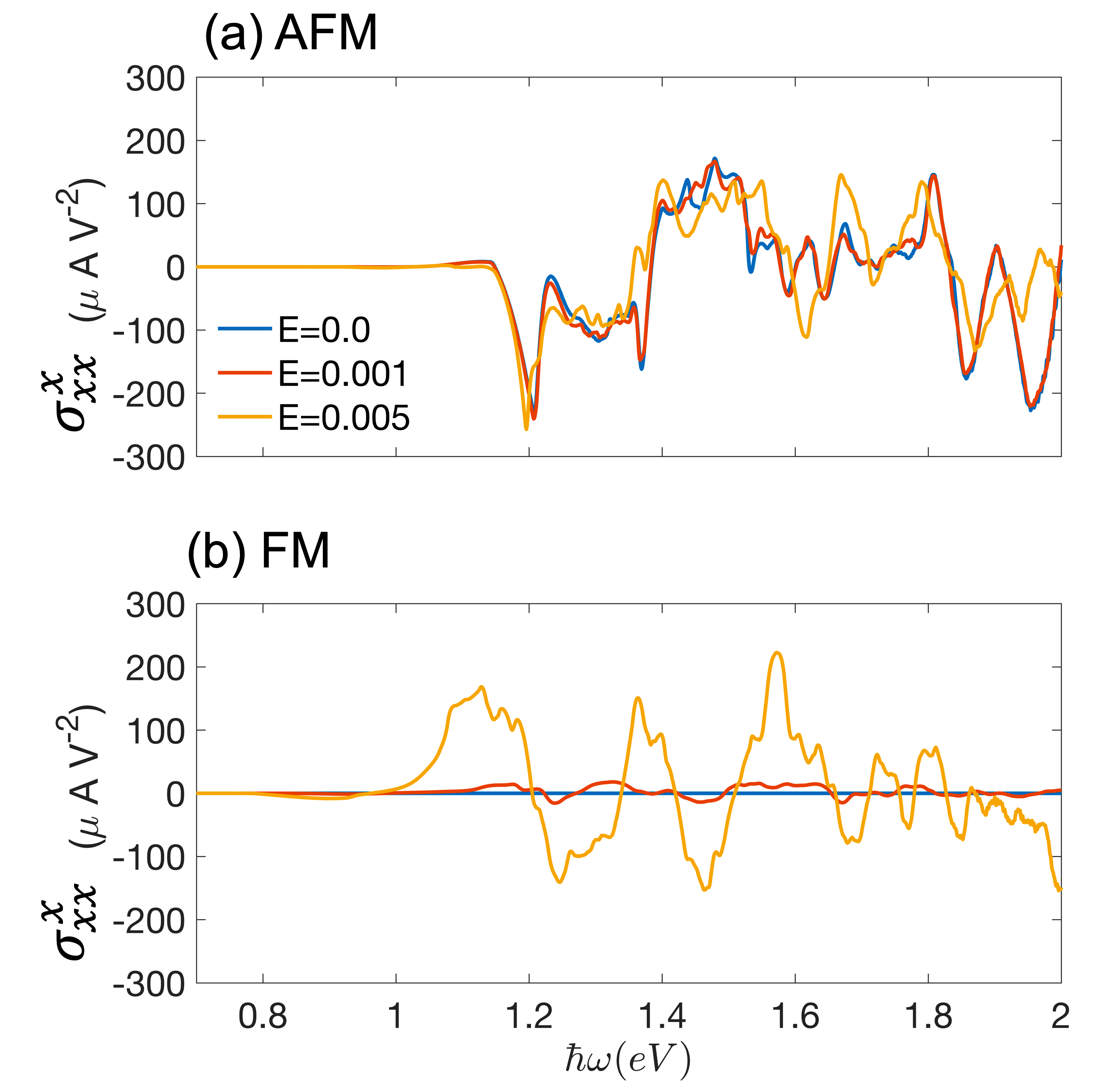}
\end{center}
\caption{The photoconductivity $\sigma^x_{xx}$ in different electric field applied for the (a) AFM and (b) FM phases. The electric field $E$ is in unit of V/$a.u.$ ( 1 $a.u. = 0.53$ \AA).}
\label{SM-Gate}
\end{figure*}

\end{document}